\documentclass[a4paper, onecolumn, 12pt]{IEEEtran}

\usepackage[cmex10]{amsmath}
\usepackage{amssymb}
\usepackage{array}
\usepackage{graphicx}
\usepackage{amssymb,amsmath}
\usepackage{cite}
\usepackage{multirow}
\usepackage{algorithm}
\usepackage{algpseudocode}
\usepackage{mathtools}

\usepackage{caption} 
\newcommand{\bor}[1]{{\color{red} #1}}

\renewcommand{\vec}[1]{\ensuremath{\boldsymbol{#1}}}

\usepackage{tikz}
\usetikzlibrary{positioning}
\usetikzlibrary{babel}
\usepackage[a4paper]{geometry}
\pgfdeclarelayer{background}
\pgfdeclarelayer{foreground}
\pgfsetlayers{background,main,foreground}

\def\remark{
  \let\go\relax
  \ifvmode\vskip-\lastskip\fi
  \noindent{\it Remark\/.}%
  \enskip\relax\ignorespaces\go}
\newcommand{\bs}[1]{\ensuremath{\boldsymbol{#1}}}

\begin{document}

\title{Optimization of NB QC-LDPC Block Codes and Their Performance Analysis }

\author{
\IEEEauthorblockN{Irina E. Bocharova$^{1,2}$, 
  Boris D. Kudryashov$^{1,2}$,
  Evgenii P. Ovsyannikov$^3$,
  Vitaly Skachek$^2$, and
  T$\ddot{\text a}$hvend Uustalu$^2$
}
\vspace{1mm}
\IEEEauthorblockA{
	\small
	\begin{tabular}{c@{\hspace{.5cm}}c}
		               &   \\
		$^1$University of Information  &$^3$State University of \\
		 Technologies, Mechanics and Optics                            & Aerospace Instrumentation \\
		St. Petersburg, 197101, Russia                      &  St. Petersburg, 190000,  Russia\\
		$^2$University of Tartu, 51009, Estonia  &  Email: {eovs@mail.ru}\\
		Email: \{irinaboc, boriskud\}@ut.ee &\\
		 \{vitaly.skachek, tahvend.uustalu\}@ut.ee &\\
	\end{tabular}
	\vspace{-4mm}
}
}

\maketitle

\begin{abstract}
We propose a novel approach for optimization of nonbinary (NB) quasi-cyclic (QC) LDPC codes. In this approach, first, the base parity-check matrices are constructed by a simulated annealing method, and then these matrices are labeled by the field elements, while maximizing the so-called generalized girth of the Tanner graph. Tightened random coding bounds, which are based on the average binary spectra of ensembles of ``almost regular'' NB LDPC codes of the finite lengths over the extensions of the binary Galois field, are derived. The FER performance of the sum-product BP decoding of ``almost regular'' NB QC-LDPC block codes is estimated experimentally, and 
is also compared  to that of the optimized  binary QC-LDPC block code in the 5G standard. The FER performance is also compared to the finite-length random coding bounds. 
It is observed that in the waterfall region, the simulated FER performance of the BP decoding is about  0.1~--~0.2 dB away from the presented finite-length bounds on the error probability of the ML decoding.
\end{abstract}

\section{Introduction}

Nonbinary (NB) LDPC block codes over arbitrary finite fields were introduced and analyzed in \cite{gallager}, where the average weight spectra for random ensembles of regular NB LDPC codes were derived. After the rediscovery of LDPC codes in the nineties, the generalized belief propagation (BP) decoding for an NB variant of LDPC codes was presented in \cite{mackay}. In that paper, it was demonstrated for the first time that the binary images of 
the NB LDPC codes over the extensions of the binary Galois field can significantly outperform the binary LDPC codes of the same rate and length. Moreover, it was shown therein that the increase in the code alphabet size yields an improvement in the code performance at the cost of larger decoding complexity. Starting with \cite{mackay}, the term ``{\em NB LDPC codes}'' is often used for the binary images of the NB LDPC codes over the extensions of the binary field. In the sequel, we use the terms ``binary images of NB LDPC codes'' and ``NB LDPC codes'' interchangeably. 

In the prior literature, a lot of attention has been paid to NB LDPC codes with two nonzero elements in each column of their parity-check matrices. Such codes were studied in \cite{hu2004binary,poulliat2008design}. An experimental comparison of binary and NB LDPC codes, including NB LDPC codes given by the parity-check matrices with more than two nonzero elements in their columns, was performed in \cite{orang}.
In those papers, the Galois extension fields GF$(2^m)$ with $2 \le m \le 10$ were considered. It was confirmed that the NB LDPC codes of short and moderate lengths outperform their binary counterparts with the same parameters. It was also noticed that the increase in the value of $m$ leads, not always monotonically, to improved performance of iterative decoding. This non-monotonicity 
was observed also in \cite{andryanova2009binary,kasai2011weight}, where decoding thresholds for ensembles of
random NB LDPC codes were studied. 

The irregular NB LDPC codes, in general, outperform their regular counterparts, similarly to the binary case. 
Nevertheless, if $m$ is large enough, say $6 \le m \le 8$, then, typically, the NB LDPC codes with two nonzero elements in each column provide for better error performance to date. However, if $m \leq 4$, then the NB LDPC codes with a larger number of nonzero elements in each column are preferable. For example, in \cite{mackay}, it was shown that for $m=4$, the best results are obtained when the average column weight is $2.4$. Moreover, for smaller field sizes, the larger column weights provide for better performance.
In \cite{li-fair}, the degree distribution of long NB LDPC codes was optimized for NB LDPC codes over GF$(3)$ and GF$(4)$. In the same paper, the BP decoding thresholds for the rate $1/2$ NB LDPC codes over GF$(2^m)$, $2 \le m \le 6$, were obtained. The authors conclude that if the column weight is larger than 3, the thresholds for the small field sizes are better than for the large counterparts, while for the column weight smaller than $2.5$, the performance improves with the increase in~$m$. 

Optimization of practical NB LDPC codes and simplification of their decoding algorithms were extensively studied in the literature (see, for example, \cite{chang2012}, \cite{hareedy2018combinatorial}, \cite{Boch2016} and the references therein). Popular families of codes in this context are QC and photograph-based NB LDPC codes.  However, the prior research mainly focused on the NB LDPC codes with two nonzero elements in each column with $m \ge 6$.  

In the current paper,  we propose a novel approach for optimization of NB QC LDPC block 
codes over GF$(2^m)$ for $m\le 6$. In this approach, the simulated annealing technique (see \cite{delahaye2019simulated} and the references therein) is applied to optimization of the code base matrix. Then, the optimization of the degree matrix is performed by the algorithm in~\cite{Boch2016}. Finally, the labeling of the resulting degree matrix by the elements of the field GF$(2^m)$ in a way that maximizes the {\em generalized girth} of the corresponding Tanner graph is carried out. This last step is equivalent to satisfying the full-rank condition in \cite{poulliat2006using}.  
                   
The FER performance of the generalized BP decoding of the NB QC LDPC codes, which are constructed in this work, is compared to the Shannon lower bound \cite{Shannon1959} and the Poltyrev upper tangential sphere (TS) bound \cite{poltyrev1994bounds} on the error probability of the maximum-likelihood (ML) decoding. The upper bound in \cite{poltyrev1994bounds} requires knowledge of the weight spectrum of the code. One of the approaches to exploitation of this bound is based on 
the average code weight spectra for code ensembles. For the ensembles of binary regular LDPC codes and LDPC codes over arbitrary nonbinary fields, the
average weight spectra were derived in \cite{gallager}. A detailed analysis of the asymptotic weight spectrum of the ensemble of the NB protograph-based LDPC codes, as well as of the NB protograph-based LDPC codes over extensions of the binary field can be found in \cite{dolecek2014non}. Ensembles of irregular NB
LDPC codes over extensions of the binary field were analyzed in \cite{kasai2011weight}. Estimates on the thresholds of the ML decoding over an AWGN channel for ensembles of the NB LDPC codes over the extension of the binary field were also presented in \cite{dolecek2014non}. 
Finite length upper bounds on the error probability of the ML decoding over the AWGN channel obtained by using precise
average weight enumerators for both binary random regular LDPC codes and for random regular NB LDPC
codes over GF$(2^m)$ along with the asymptotic ML decoding thresholds were derived in \cite{bocharova2017FL}. 

In the current paper, by using technique in \cite{bocharova2017} for computing precise average spectra for ensembles of LDPC codes, we derive a tighter bound on the error probability of the ML decoding for the ensemble of ``almost regular'' NB  LDPC codes over GF$(2^m)$. 
This bound allows for the analysis of the random codes with degree distributions which mimics the degree distribution 
of the practical NB LDPC codes designed using the simulated annealing technique.
 
The main contributions of the paper are as follows:
\begin{itemize}
\item A new random ensemble of almost regular NB LDPC codes is introduced and analyzed; 
\item Finite length random coding bounds for ensembles of almost regular NB LDPC codes are derived;
\item Simulated annealing based approach for finding good base matrices of the NB LDPC codes is suggested. 
\end{itemize}

%
%
%
%
%

This paper is organized  as follows. In Section~\ref{prelim}, the necessary definitions and background are given. We describe the proposed optimization technique in Section~{\ref{Annealing}}. In Section~\ref{AlmostRegular}, we describe a new ensemble of ``almost regular'' NB LDPC codes over extensions of the  
binary field. In the same section, we derive a formula for the average binary weight spectrum of the proposed ensemble. The simulation results and their comparison with the tightened bounds are performed in Section \ref{Simul}. The paper is concluded by a short discussion. For completeness, known bounds on the error probability of the ML decoding are presented in the Appendix.  

\section{Preliminaries \label{prelim}}

A  rate $R=b/c$  NB QC-LDPC code over the extension field GF($q$), $q = 2^m$, $m \ge 2$, is defined by its polynomial parity-check matrix of size $(c-b)\times c$ 
\[
H(D) = \{ h_{ij}(D) \} \;,
\] 
where $h_{ij}(D)$ are polynomials of the formal variable $D$ with the coefficients in GF($q$). 
In the sequel, $h_{ij}(D)$ are either zeros or monomials, and 
\[
H(D) = \{ \alpha_{ij}D^{w_{ij}} \} \;, w_{ij}\in \{0,1,...,\nu\}\;,  \alpha_{ij} \in {\rm GF}(2^m), \; i=1,...,c-b, \; j=1,...,c,
\] 
where $\nu$ denotes the maximal degree of a monomial.  
The corresponding $q$-ary parity-check matrix of the $[Lc,Lb]$ NB QC-LDPC block code  is obtained by replacing $D^{w_{ij}}$ by the $w_{ij}$-th power of a circulant permutation matrix of order $L$. The parameter $L$ is called the {\em lifting factor}. 
The parity-check matrix in the binary form is obtained by replacing the non-zero elements of the $q$-ary parity-check matrix by the binary $m\times m$ matrices, which are the companion matrices of the corresponding field elements \cite{macwilliams1977theory}.

Let $\vec \alpha_i = (\alpha_{i1},\alpha_{i2},...,\alpha_{iw_i})$ be a vector consisting of nonzero elements of the
$i$-th row of $H(D)$, and let $w_i$ be the number of nonzero elements of that row. 
After replacing these nonzero elements by their binary $m\times m$ companion matrices, we obtain an 
$m\times mw_i$ parity-check matrix of a linear code which we call the $i$-th {\em constituent} code of the NB LDPC code.

To facilitate the low encoding complexity, we consider the
parity-check matrices having the form (see, for example, \cite{Boch2016}):
 \begin{equation} \label{bidiag}
H(D)=
\begin{pmatrix}
 H_{\rm inf}(D) & \vec h_0(D)& H_{\rm bd}(D) 
\end{pmatrix},
\end{equation}
where  $H_{\rm bd}(D)$ is a bidiagonal matrix of size $(c-b)\times (c-b-1)$,
$\vec h_0(D)$ is a column with two 
nonzero elements, 
and $H_{\rm inf}(D)$ is an arbitrary monomial submatrix of the proper size.
The submatrix $H_{\rm inf}(D)$ corresponds to the information part of a codeword.
A binary matrix $B=\{b_{ij}\}$ of the same size as $H(D)$ is called the {\em base} matrix of $H(D)$ if 
$b_{ij}=1 \Longleftrightarrow h_{ij}(D)\neq 0$.

In the process of searching for the optimized parity-check matrices, we represent $H(D)$ by the following two matrices: the degree matrix $H_w=\{ w_{ij} \}$ and the matrix of the field coefficients $H_c=\{ \alpha_{ij} \}$, which are obtained by labeling the nonzero elements of $B$ by the monomial degrees and by the nonzero field elements, respectively. 
In $H_w$ and $H_c$ we write ``$-1$'' in the positions corresponding  to the zero elements of $B$.   

If the matrices $B$ and $H(D)$ have $J$ nonzero elements in each column and $K$ nonzero elements in each row, we call them $(J,K)$-{\em regular} LDPC codes, otherwise the codes are called 
{\em irregular}. In the sequel, we focus on the NB QC-LDPC codes with the columns of weight two and three in their matrix $B$. We call such codes {\em almost regular}.

In order to construct the NB QC-LDPC codes, we begin by finding good matrices $B$, $H_w$, $H_c$. 
First, we onptimize the the base matrix $B$. 
 In the next section, we explain the proposed approach to optimization of the base matrices with a given degree distribution by using the simulated annealing technique.

Next, we present some notions from the field of graph theory. A (simple) graph $\mathcal{G}$ is determined by a set of \textit{vertices} $\mathcal{V} = \{v_i\}$ and a set of \textit{edges} $\mathcal{E} = \{e_i\}$, where each edge is a set with exactly two vertices (we say that the edge \emph{connects} those two vertices). The \textit{degree of a vertex} denotes the number of edges that are connected to it. If all vertices have the same degree $l$, the \textit{degree of the graph} is $l$, or, in other words, the graph is $l$-\textit{regular}.

Consider the set of vertices $\mathcal{V}$ of a graph partitioned into $t$ disjoint subsets $\mathcal{V}_k$, $k=0,1,\ldots,t-1$, where $\mathcal{V} = \cup_{k=0}^{t-1} \mathcal{V}_k$. Such a graph is said to be $t$-\textit{partite} if no edge connects two vertices from the same set $\mathcal{V}_k$, $k=0,1,\ldots,t-1$.


A \textit{walk} of length $N$ in a graph is an alternating sequence of $N+1$ vertices $v_i$, $i=1,2,\ldots,N+1$, and $N$ edges $e_i$, $i=1,2,\ldots,N$, with $e_i \neq e_{i+1}$. If $v_1 = v_{N+1}$ then the walk is a \textit{cycle}.
A cycle is called \textit{simple} if all its vertices and edges are distinct, except for the first and last vertex, which coincide. The length of the shortest simple cycle is called the \textit{girth} of the graph.

A parity-check matrix $H$ of a rate $R=k/n$ LDPC block code can be interpreted as the \textit{biadjacency matrix}\cite{Asratian1998} of a bipartite graph, the so-called \textit{Tanner graph} \cite{Tanner1981}, having two disjoint subsets $\mathcal{V}_0$ and $\mathcal{V}_1$ containing $n$ and $n-k$ vertices, respectively. The $n$ vertices in $\mathcal{V}_0$ are called the \textit{symbol or variable nodes}, while the $n-k$ vertices in $\mathcal{V}_1$ are called the \textit{constraint or check nodes}. If the underlying LDPC block code is $(J, K)$-regular, the symbol and constraint nodes have degrees $J$ and $K$, respectively.

A parity-check matrix $H$, whose columns have weight two, can itself be considered the incidence matrix of a graph.  In this graph, the vertices correspond to the rows of $H$ and the edges correspond to its columns. The corresponding Tanner graph consists of the vertices of two types corresponding to the rows and columns of $H$, and its edges correspond to the nonzero elements. Consequently, the girth of the Tanner graph is two times larger than the girth of the graph with $H$ being its incidence matrix.     

A \textit {hypergraph} is a generalization of a graph in which the \textit{ hyperedges}
are the sets of vertices of arbitrary sizes larger than or equal to two. 
A parity-check matrix $H$ of a rate $R=k/n$ LDPC block code can be interpreted as the incidence matrix of a hypergraph. 
A hypergraph is called \textit{$s$-uniform} if every hyperedge connects (contains) $s$
vertices. The \textit{degree  of
a vertex} in a hypergraph is the number of
hyperedges that are connected to it.
If all the vertices have the same degree then it
is the \textit {degree of the hypergraph}.
The hypergraph is \textit {$l$-regular} if every vertex
has the same degree $l$.

Let the set $\mathcal V$ of vertices of an $s$-uniform hypergraph
be partitioned into $t$ disjoint subsets $\mathcal V_{j}$, $j=1,2,\dots,t$, where $\mathcal{V} = \cup_{k=0}^{t-1} \mathcal{V}_k$.
A hypergraph is said to be \textit{$t$-partite} if no edge contains two vertices from the same set $\mathcal V_{j}$, $j=1,2,\dots,t$.

\section{Simulated annealing technique for constructing base matrices with column weight two and three
\label{Annealing}}

When constructing the NB LDPC codes over relatively small fields GF$(2^m)$, $m<6$, both the density evolution analysis and the simulations show that the average column weight of the base matrix should be close to the interval [2.2, 2.4]. This motivates the choice for the structure of the base matrix in this work. In this work, we consider the base matrices with columns of weight two and three, which additionally have the aforementioned bi-diagonal structure. 

The base matrices with the column weight two can be viewed as incidence matrices of graphs. 
Their rows correspond to vertices, and the columns correspond to the edges of these graphs. 
For the base matrices with the column weight larger than two, the columns can be interpreted as 
hyperedges. Such base matrices represent 
incidence matrices of hypergraphs (see, \cite{schmidt2003expander, Boch2016}).

The problem of constructing graphs with a given girth is well-studied in the graph theory. Its counterpart for constructing hypergraphs is much more involved, and it is studied to a lesser degree. The idea behind the proposed approach is as follows. 
We first construct a graph whose incidence matrix determines a higher rate base LDPC code. Then, by ``gluing'' the edges of the graph, we construct a hypergraph whose incidence matrix determines a base LDPC code with the required parameters. When searching for the graph, we aim at maximizing the girth of the corresponding Tanner graph. During this process, we aim at finding a base LDPC code with Tanner graph having the same girth as the girth of the initial graph. 
 
Next, we discuss how the hypergraph can be obtained from a given graph. Then, we explain the specifics of the simulated annealing technique, and apply it to the both steps: searching for optimized  graphs and searching for the optimal ``gluing''.

\subsection{Constructing hypergraphs from graphs \label{hyper}}

The last $(c-b)$ columns of the base matrix  $B=H(D)|_{D=1}$ in the form (\ref{bidiag}) 
correspond to a cycle passing through all nodes of the corresponding graph (or hypergraph) exactly once. 
Such a cycle is called {\em Hamiltonian cycle}. In the sequel, we search for 
good graphs and hypergraphs with Hamiltonian cycle.

\begin{figure}
  \centering
  \begin{tikzpicture}
[scale=1.0, transform shape,
symb/.style={circle,draw=black!50,fill=black!,thick,radius=1.5mm},
check/.style={circle,draw=black!50,fill=white!,thick,radius=1.5mm},
]
%
%
%
%
%
%

\small

\node [check] (a1) at ( 180:2cm) {};    \node  ()  at ( 180:2.5cm) {\large $a$} ;
\node [check] (a2) at ( 120:2cm) {};  \node  ()  at (  120:2.5cm) {\large $b$} ;
\node [check] (a3) at ( 60:2cm) {};     \node  ()  at (  60:2.5cm) {\large $c$} ;
\node [check] (a4) at ( 0:2cm) {};     \node  ()  at (  0:2.5cm) {\large $d$} ;
\node [check] (a5) at ( -60:2cm) {};    \node  ()  at (  -60:2.5cm) {\large $e$} ;
\node [check] (a6) at ( -120:2cm) {};  \node  ()  at (  -120:2.6cm) {\large $f$} ;

\draw (a1) to node [left] {1} (a2);
\draw (a2) to node [above] {2} (a3);
\draw (a3) to node [right] {3} (a4);
\draw (a4) to node [right] {4} (a5);
\draw (a5) to node [below] {5} (a6);
\draw (a1) to node [left] {6} (a6);
\draw [very thick] (a1) to node [left,xshift=-7mm,yshift=-3mm] {7} (a3);
\draw [very thick, dashed] (a2) to node [left,xshift=-6mm,yshift=6mm] {8} (a4);
\draw [very thick] (a3) to node [left, xshift=4mm,yshift=1.1cm] {9} (a5);
\draw [very thick, dashed] (a4) to node [left,xshift=-8mm,yshift=-7mm] {10} (a6);
\draw (a1) to node [below,xshift=-5mm,yshift=1mm] {11} (a4);
\draw (a2) to node [left,xshift=1mm,yshift=5mm] {12} (a5);
\draw (a3) to node [left,xshift=1mm,yshift=-6mm] {13} (a6);

\node () at (0,-5){
$H=
\left(
\begin{array}{ccccccccccccc|r}
1&2& 3&4&5 & 6 &7& 8& 9&10&11&12 & 13&\\ \hline  
1& &  &   &  & 1&\text{\textcircled{1}} &  &   &        & {1}  &   &    &a  \\
1&1& &   &  &   & &\boxed{1} &  &     &   &1 & &b  \\
 &1&1&   &  &   &\text{\textcircled{1}} &  &\text{\textcircled{1}} & & &&1&  c   \\ 
 & &1 &1 &   &  & & \boxed{1}& &\boxed{1}       & 1& &  &d  \\
 & &  &1 &1 &   & &  &\text{\textcircled{1}} &    &    &1 &  &e \\
 & &  &  &1 & 1 & &  & &\boxed{1}   &    & & 1&f  \\
\end{array}
\right)
$
};

%
%
%
%
%
%
%
%

\end{tikzpicture}
  \caption{ 
  Constructing hypergraph from graph.
    \label{h1}}
\end{figure}
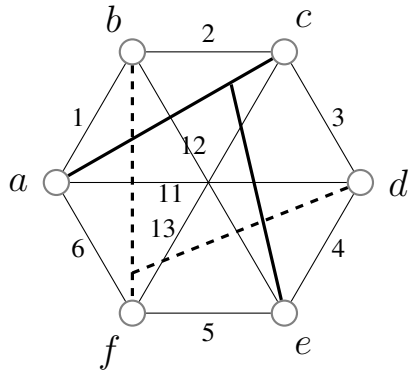

\begin{figure}
  \centering
  \begin{tikzpicture}
[scale=1.0, transform shape,
symb/.style={circle,draw=black!50,fill=black!,thick,radius=1.5mm},
check/.style={circle,draw=black!50,fill=white!,thick,radius=1.5mm},
]
%
%
%
%
%
%
\small
\node [check] (a1) at ( 180:2cm) {};    \node  ()  at ( 180:2.5cm) {\large $a$} ;
\node [check] (a2) at ( 120:2cm) {};  \node  ()  at (  120:2.5cm) {\large $b$} ;
\node [check] (a3) at ( 60:2cm) {};     \node  ()  at (  60:2.5cm) {\large $c$} ;
\node [check] (a4) at ( 0:2cm) {};     \node  ()  at (  0:2.5cm) {\large $d$} ;
\node [check] (a5) at ( -60:2cm) {};    \node  ()  at (  -60:2.5cm) {\large $e$} ;
\node [check] (a6) at ( -120:2cm) {};  \node  ()  at (  -120:2.6cm) {\large $f$} ;

\draw (a1) to node [left] {1} (a2);
\draw (a2) to node [above] {2} (a3);
\draw (a3) to node [right] {3} (a4);
\draw (a4) to node [right] {4} (a5);
\draw (a5) to node [below] {5} (a6);
\draw (a1) to node [left] {6} (a6);


\draw (a1) to node [below,xshift=-5mm,yshift=1mm] {11} (a4);
\draw (a2) to node [left,xshift=1mm,yshift=5mm] {12} (a5);
\draw (a3) to node [left,xshift=-3mm,yshift=-6mm] {13} (a6);

\node () at (0,-5){
$H=
\left(
\begin{array}{ccccccccccc|r}
1&2& 3&4&5 & 6 &(7,9)& (8,10)& 11&12 & 13&\\ \hline  
1& &  &   &  & 1&\text{\textcircled{1}} &         & {1}  &   &    &a  \\
1&1& &   &  &   & &\boxed{1} &  &        1 & &b  \\
 &1&1&   &  &   &\text{\textcircled{1}} &  & &&1&  c   \\ 
 & &1 &1 &   &  & & \boxed{1}&      1& &  &d  \\
 & &  &1 &1 &    &\text{\textcircled{1}} &    &    &1 &  &e \\
 & &  &  &1 & 1   & &\boxed{1}   &    & & 1&f  \\
\end{array}
\right)
$
};

\draw [very thick] (a1) to node [left,xshift=-7mm,yshift=-3mm] {} (a3);
\draw [very thick] (a5) to node [left,xshift=-7mm,yshift=-3mm] {} (0.3, 1.3);

\draw [very thick, dashed] (a2) to node [left,xshift=-7mm,yshift=-3mm] {} (a6);
\draw [very thick, dashed] (a4) to node [left,xshift=-7mm,yshift=-3mm] {} (-1, -1.2);

%
%
%
%
%
%
%
%

\end{tikzpicture}
  \caption{ 
  Constructing hypergraph from graph.
    \label{h2}}
\end{figure}

When searching for the base hypergraphs, as a search criteria, we employ the girth of the corresponding Tanner 
graph and approximate cycle extrinsic message degree (ACE) \cite{vukobratovic2008generalized}.
We note that the exhaustive search for hypergraphs with good parameters is computationally infeasible. 
In order to reduce the search space, we split the process into two steps. 

To find a hypergraph with $n_c$ check nodes and $n_v$ variable nodes, we
first construct a graph with  $n_c$ vertices and $n_v+2c_3$ edges with good parameters, 
where $c_3$ is the number of hyperedges containg three vertices. 
Next, we obtain $c_3$ hyperedges by converting $c_3$ pairs of edges with a joint check node into 
hyperedges.   
This method is illustrated in Figs \ref{h1}, \ref{h2}.
In Fig. \ref{h1} the graph with the girth equal to 3 is shown. 
Its incidence matrix represents a base matrix whose Tanner graph has girth  6. 
One hyperedge can be obtained from edges 7 and 9 (shown by bold lines), another 
hyperedge can be obtained from edges 8 and 10 (shown by dashed lines). 
The corresponding elements of the parity-check matrix are marked by circles and squares,
respectively.  

In Fig.~\ref{h2}, we show a hypergraph with two hyperedges and the corresponding 
base matrix with two columns of weight 3. It is easy to verify that the girth of the Tanner graph of the resulting base matrix is equal to 6, 
therefore the girth of the initial graph is preserved (which cannot be guaranteed in the general case).  
 
\bor{
}

We apply simulating annealing to construct base matrices of size $r_{\rm b}\times (c_2+c_3)$ such that all of the following conditions hold:
\begin{itemize}
\item there are $c_2$ columns with Hamming weight two;
\item there are $c_3$ columns with Hamming weight three;
\item the girth of the corresponding Tanner graph is maximized;
\item for matrices with equal girth, the number of shortest cycles is minimized.
\end{itemize}

Equivalently, the goal is to construct a hypergraph  with $r_{\rm b}$ vertices and $c_2 + c_3$
hyperedges such that all of the following conditions hold:
\begin{itemize}
\item there are $c_2$ hyperedges connecting two vertices;
\item there are $c_3$ hyperedges connecting three vertices;
\item the girth of the Tanner graph corresponding to that hypergraph is maximized;
\item among hypergraphs with equal girth of their Tanner graphs, the number of shortest cycles is minimized;
\item there exists a Hamiltonian cycle, composed of only hyperedges connecting two vertices.
\end{itemize}

\begin{figure}[h]
  \centering
 \begin{tikzpicture}[node distance=10mm,
     v/.style={minimum size=8mm, circle, draw, fill=white},
   ]
    \node [v] (v1) {$v$};
    \node [v] (v2) [above right=0.5cm and 1cm of v1] {$u$};
    \node [v] (v3) [below right=0.5cm and 1cm of v1] {$w$};
    \draw (v1) -- (v2);
    \draw (v1) -- (v3);
    \node (e1a) [above = 0.3cm of v1] {};
    \node (e1b) [below = 0.3cm of v1] {};
    \node (e1c) [left = 0.3cm of v1] {};
    \draw (v1) -- (e1a);
    \draw (v1) -- (e1b);
    \draw (v1) -- (e1c);
    \node (e2a) [above left = 0.3cm of v2] {};
    \node (e2b) [below right = 0.3cm of v2] {};
    \node (e2c) [above right = 0.3cm of v2] {};
    \draw (v2) -- (e2a);
    \draw (v2) -- (e2b);
    \draw (v2) -- (e2c);
    \node (e3a) [below right = 0.3cm of v3] {};
    \node (e3b) [below left = 0.3cm of v3] {};
    \node (e3c) [above right = 0.3cm of v3] {};
    \draw (v3) -- (e3a);
    \draw (v3) -- (e3b);
    \draw (v3) -- (e3c);

    \node [v] (w1) [right =3cm of v1] {$v$};
    \node [v] (w2) [above right=0.5cm and 1cm of w1] {$u$};
    \node [v] (w3) [below right=0.5cm and 1cm of w1] {$w$};
    \begin{pgfonlayer}{background}
      \path [fill=gray] (w1.center) -- (w2.center) -- (w3.center) -- (w1.center);
   \end{pgfonlayer}
    \node (f1a) [above = 0.3cm of w1] {};
    \node (f1b) [below = 0.3cm of w1] {};
    \node (f1c) [left = 0.3cm of w1] {};
    \draw (w1) -- (f1a);
    \draw (w1) -- (f1b);
    \draw (w1) -- (f1c);
    \node (f2a) [above left = 0.3cm of w2] {};
    \node (f2b) [below right = 0.3cm of w2] {};
    \node (f2c) [above right = 0.3cm of w2] {};
    \draw (w2) -- (f2a);
    \draw (w2) -- (f2b);
    \draw (w2) -- (f2c);
    \node (f3a) [below right = 0.3cm of w3] {};
    \node (f3b) [below left = 0.3cm of w3] {};
    \node (f3c) [above right = 0.3cm of w3] {};
    \draw (w3) -- (f3a);
    \draw (w3) -- (f3b);
    \draw (w3) -- (f3c);
  \end{tikzpicture}
  \caption{Merging two edges $\{u,v\}$ and $\{v,w\}$ to form a hyperedge $\{u,v,w\}$.}
  \label{merging}
\end{figure}
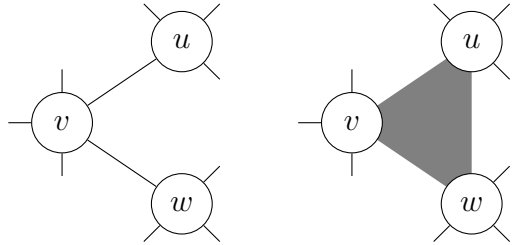

Consider any pair of incident edges $\{u,v\}$ and $\{v,w\}$ in a graph. We replace these two edges by a
hyperedge $\{u,v,w\}$ as in Fig.~\ref{merging}. We call this operation \emph{merging} of the two edges.

Simulated annealing is an optimization technique which allows to escape local extrema. At each iteration of the optimization algorithm, the current solution and a new solution are compared by using an objective function. Improved solutions are always chosen, while a fraction of non-improved solutions are chosen with a probability depending on the so-called \emph{temperature parameter}. The non-improved solutions are probabilistically chosen in order to escape possible local extrema in the search for the global extremum. 

The temperature parameter is typically non-increasing with the iterations. The terms ``energy (objective) function'', ``temperature profile'', etc., stem from the fact that this algorithm mimics a process of bringing metal to a very high temperature until `'melting'' of its structure, and then cooling it according to a very particular temperature decreasing scheme in order to reach a solid state of the minimum energy. For the detailed overview of the algorithm see \cite{delahaye2019simulated} and the references therein.

A generic description of simulated annealing is  presented in a form of pseudo-code in Fig.~\ref{Anneal}.
This algorithm finds applications in different areas, including search for good LDPC codes. 
In particular, in~\cite{usatyuk2018simulated} and \cite{sariduman2019construction}, the simulated annealing technique is used for labeling of the base matrices of the QC-LDPC codes. In~\cite{cui2014dynamic}, the simulated annealing is used for decoding of LDPC codes with dynamic schedule.

\begin{figure}
\begin{algorithmic}[0]
\Statex {\bf Choose:} An energy function $E(\cdot)$, a randomized perturbation function $p(\cdot)$, 
a number of iterations $I_{\max}$,
and  a ``temperature profile'' $\vec t =(t_1,t_2,...t_{I_{\max}})$.
\Statex {\bf Initialization:} choose a random point $\vec v_0$ in a space $V$  
\For {$I=1$ to $I_{\max}$ }\\
{
\hspace{5mm}Compute ${\vec v}_I ^{'}=p({\vec v}_I)$ 
{
\If { 
$E({\vec v}_I^{'})\le E({\vec v}_I)$} \\
\hspace{10mm} set $ {\vec v}_{I+1}={\vec v}_{I}^{'}$
\Else{ \\
\hspace{10mm} set ${ \vec v}_{I+1}={\vec v}_{I}^{'}$ with probability $\Pr=\exp \frac{E({\vec v}_I^{'})- E({\vec v}_I)}{t_I}$}
\EndIf
}
}
\EndFor
\end{algorithmic}
\caption{Simulated annealing algorithm \label{Anneal}}
\end{figure}

We construct the hypergraph in two steps, as follows:

\begin{itemize}
\setlength{\itemindent}{1cm}
\item [\bf Step 1.] Build a graph with $r_{\rm b}$ vertices and $c_2 + 2c_3$ edges, containing a fixed Hamiltonian
  cycle;
\item [\bf Step 2.] Pick $c_3$ pairwise disjoint pairs of incident edges, none of them belonging to the Hamiltonian 
  cycle, and merge them pairwise together.
\end{itemize}
Both steps are done using simulated annealing.

Next, we describe the parameters $V$, $E(\cdot)$, $p(\cdot)$ and $t$, that is, the search space, energy function, perturbation function and temperature profile, in the context of the both steps of the algorithms. For a (hyper) graph $G$ and a natural number $g$, define $N_{G, g}$ to be the number
of cycles of length $g$ in the corresponding Tanner graph. For a graph $G$ and a natural number $g$, define
$m_{G, g}$ as follows:
\begin{itemize}
\item Consider all pairs of incident edges $\{u,v\}$ and $\{v,w\}$ in $G$.
\item For each such pair, find the shortest walk between $u$ and $w$ that does not visit $v$,
   i.e. find the shortest cycle through $\{u,v,w\}$ that contains the edges $\{u,v\}$ and $\{v,w\}$.
\item Define $m_{G, g}$ to be the number of such pairs ($\{u,v\}$, $\{v,w\}$)  for which the length of the shortest 
cycle in the Tanner graph  obtained after merging edges is $g$.
\end{itemize}
Then, the simulated annealing parameters for Step 1 are:
\begin{itemize}
\item The search space $V$ is a set of all graphs with a given number of vertices and edges, containing
  the fixed Hamiltonian cycle.
\item The energy function 
\begin{equation}
E(G) = \sum_g \left(N_{G, g} + \gamma m_{G, g} \right) x^{2(g-2)}\;, 
\label{eq:energy_function_1}
\end{equation}
where $0 < x < 1$, $x$ is small. The constant $\gamma$ is picked
  experimentally. If the graph $G$ is ``illegal'', i.e. contains self-loops or parallel edges,
  then $E(G) = \infty$.
\item The perturbation function $p(G)$ maps a graph to a graph: pick a random non-cycle edge $\{u,v\}$; delete $\{u,v\}$.
  With probability $1/2$ swap $u$ and $v$. Pick a random vertex $w$ such that $\{u,w\}$ is not
  in the graph, and add $\{u,w\}$ to the graph.
\item The temperature profile: $t_I = t_0 \cdot t_{\mathrm{step}}^I$, $I=1,2,...,I_{\max}$,
  where \[ t_{\mathrm{step}} = \left( \frac{t_{I_{\max}}}{t_0} \right)^{\frac{1}{I_{\mathrm{max}}}}.\]
  The constants $t_0$ and $t_{I_{\max}}$ are picked experimentally.    
\end{itemize}
The parameters for Step 2 are:
\begin{itemize}
\item The search space $V$: space of all possible sets $S$ of pairwise-disjoint pairs to merge;
\item The energy function 
\begin{equation}
E(S) = \sum_g N_{G, g} x^{2(g-2)}\;,
\label{eq:energy_function_2} 
\end{equation}
where $0 < x < 1$, $x$ is small.
  Here $G$ is the hypergraph obtained by merging the pairs of edges in $S$. If $S$ contains
  some edges in multiple pairs or is otherwise ``illegal'', then $E(S) = \infty$ instead.
\item The perturbation function $p(S)$ maps the set $S$ to the set of $c_3$ pairs of edges to be merged: pick a random element of $S$ and delete it. Pick a
  random vertex $u$ and two neighbors $v$ and $w$. If neither $\{u,v\}$ nor $\{u,w\}$ are cycle edges,
  add the pair $(\{u,v\},\{u,w\})$ to $S$; otherwise repeat this step.
\item The temperature profile:  $t_I = t_0 \cdot t_{\mathrm{step}}^I$, $I=1,2,...,I_{\max}$
  where \[ t_{\mathrm{step}} = \left( \frac{t_{I_{\max}}}{t_0} \right)^{\frac{1}{I_{\mathrm{max}}}}.\]
  The constants $t_0$ and
  $t_{I_{\max}}$ are picked experimentally.    
\end{itemize}

All the randomly chosen parameters use the uniform distribution in the corresponding ranges. The other parameters in our experiments are chosen as
follows: $I_{\mathrm{max}} = 10^6$, $x = 0.1$, and $\gamma = 20$. The values of the parameters 
$t_0$ and $t_{I_{\max}}$ for the two optimization steps are presented in Table~\ref{SA}.

\begin{table}
\arraycolsep=20pt
\def\arraystretch{1.5}
\setlength{\tabcolsep}{5mm}

\center
\begin{tabular}{|c|c|c|c|c|} 
 \hline
        & \multicolumn{2}{c|}{Step 1} & \multicolumn{2}{c|}{Step 2} \\ \cline{2-5}
  $c_3$ & $t_0$ & $t_{I_{\max}}$ & $t_0$ & $t_{I_{\max}}$ \\
  \hline
  $20$ & $1.603$ & $10^{-7}$ & $10$ & $10^{-7}$ \\
  $15$ & $1.375$ & $10^{-7}$ & $10$ & $10^{-7}$ \\
  $10$ & $1.298$ & $10^{-7}$ & $10$ & $10^{-7}$ \\
  $0$ & $0.862$ & $10^{-7}$ & --- & --- \\ \hline
  
\end{tabular}
\caption{Example set of parameters of simulated annealing \label{SA} for matrices of size $26 \times 52$}
\end{table}


Finally, we explain the technique for fast computation of the the energy function. 
Since we use a large number of iterations in the simulated annealing, it is important
to calculate the energy functions efficiently. For that reason, in the energy functions~(\ref{eq:energy_function_1}) and~(\ref{eq:energy_function_2}), the
values $N_{G, g}$ are approximated by using the dynamic programming, as it is explained below.
In the following, we consider the four-tuples $(\ell, u, e, v)$, where $\ell$ is a positive integer,
$u$ is a vertex, $e$ is a (hyper-) edge and $v$ is a vertex contained in $e$. Denote by
$\mathrm{dp}(\ell, u, e, v)$ the number of walks in the graph $G$ with the following properties:
\begin{itemize}
  \item the length of the walk is $\ell$;
  \item the first vertex is $u$;
  \item the last (hyper-)edge is $e$;
  \item the last vertex is $v$;
  \item the walk never visits any (hyper-)edge twice in a row.
\end{itemize}
Clearly, if $\ell > 1$, then
\begin{equation} \label{dp_trans}
    \mathrm{dp}(\ell, u, e, v) = \sum_{w\in f, \, w \in e \; : \; f\neq e, \, w\neq v} \mathrm{dp}(\ell - 1, u, f, w) \; .
\end{equation}

Here, the sum is taken over all  $w$, $w \ne v$, that are contained in $e$, and
all $f$, $f \ne e$,  that are incident with $w$. We have
\begin{equation} \label{dp_base}
    \mathrm{dp}(1, u, e, v) =
    \begin{cases}
      1, & \text{ if } u, v \in e \text{ and } u \ne v; \\
      0 & \text{ in all other cases}.
      \end{cases}
\end{equation}
  By using equations (\ref{dp_trans}) and (\ref{dp_base}), we compute $\mathrm{dp}(\ell, u, e, v)$
  for all four-tuples $(\ell, u, e, v)$. Then,
  \[N_{G, g} \triangleq \sum_{u, e} \mathrm{dp} (g, u, e, u).\]

  This way, the precise number of the shortest cycles is computed. For the larger values
  of $g$, additionally, the algorithm counts closed walks that are not simple cycles.
  Moreover, the number of cycles of length $g$ is multiplied by $2g$, because each
  cycle of length $g$ has $g$ possible starting points and 2 possible traversal directions.

  These inaccuracies do not negatively affect the solutions found by the simulated annealing,
  as the energy function is still mostly ``monotone'': better solutions have lower
  energy functions, and the number of the shortest cycles (which is exact) always dominates
  the energy function.

\subsection{Constructing the degree matrix and matrix of coefficients}

Optimization of the monomial parity-check matrix $H(D)=\{ \alpha_{ij} D^{w_{ij}} \}$, $i=1,...,c-b,j=1,...,c$,
contains optimization of the base matrix $B$, the degree matrix $H_w=\{w_{ij}\}$, and the matrix of 
coefficients $H_c=\{\alpha_{ij}\}$. 

The matrix $B$ is selected by the simulated annealing.
Optimization of the degree matrix is performed by using the same techniques, which are
used for constructing binary QC LDPC codes, for example, by using the algorithm suggested 
in \cite{Boch2016}.
 
In order to construct the matrix of the coefficients $H_c$, we use the following approach. Consider a binary image of a $(J,K)$-regular $[Lc, Lb]$ NB QC LDPC code  over GF$(2^m)$, $m=4$. It is easy to see that the constituent $[Km,(K-1)m]$ codes are high-rate codes, which cannot have the minimum distance larger than two.  We search for the constituent codes with the minimal number of weight-two codewords and store a set of such codes. By assuming that we have a collection of such codes, we try to select them in such a way that the overall code performance is 
optimized. 

As a search criterion we use a combinatorial {characteristic} of NB LDPC codes. We call it a 
\emph{generalized girth}. We explain this notion by the following example.

Consider a cycle of  length six in the code Tanner graph. The corresponding 
fragment of the degree matrix can be reduced to the form
\[ 
\begin{pmatrix}
D^{w_1} & D^{w_2} &0\\
D^{w_3} & 0& D^{w_4} \\
0& D^{w_5} & D^{w_6}
\end{pmatrix}  \;.
\]
This matrix determines a cycle if and only if 
\begin{equation} \label{C1}
w_1+w_4+w_5-w_2-w_3-w_6=0 \; \mod L \;,
\end{equation}
where $L$ is a lifting degree of the QC code. Assume that this condition is fulfilled. By using a corresponding 
fragment, we obtain the corresponding fragment of the polynomial matrix $H(D)$ labeled by the field elements $\alpha_i$
\begin{equation} \label{C2}
\begin{pmatrix}
\alpha_1D^{w_1} & \alpha_2D^{w_2} &0\\
\alpha_3D^{w_3} & 0& \alpha_4D^{w_4} \\
0& \alpha_5D^{w_5} &\alpha_6 D^{w_6}
\end{pmatrix} \;.
\end{equation}
When assuming that all $\alpha_i$ are nonzero under the condition (\ref{C1}), the submatrix (\ref{C2}) is degenerate if and only if
\begin{equation} \label{C3}
\alpha_1 \alpha_4 \alpha_5 = \alpha_2 \alpha_3 \alpha_6\;,
\end{equation}
where the operations are performed in GF($2^m$). Notice that if (\ref{C3}) is rewritten via degrees
of the primitive element of the field, then this condition coincide with (\ref{C1}) up to the notations. Observe also that condition (\ref{C3}) is equivalent to the full-rank condition in \cite{poulliat2006using}.
Now we can formally define the generalized girth.  

Let $H_w$ and $H_c$ be the matrices defining the NB QC LDPC code. 
Consider two Tanner graphs, corresponding to two binary QC codes, one code is defined by 
$H_w$ and the lifting degree $L$, while the second code is defined by the same base matrix labeled by the
degrees of the entries of $H_c$ and having lifting degree $q=2^m$. A sequence of edges is called a nonbinary cycle
(generalized cycle) if it is a cycle in the both Tanner graphs. 
The length of a shortest nonbinary cycle is called a {\em generalized} girth of the NB LDPC code.   

When searching for the matrix of the coefficients $H_{c}$, we simultaneously maximize the generalized girth and minimize the multiplicity of short nonbinary cycles.

Next, we explain the proposed method for selecting constituent codes. Let $\{K_i\}$, $i=1,...,J$, 
be the set of row weights. Prior to searching for good NB LDPC codes, we construct lists of 
the constituent code candidates. The number of lists is equal to the number of different values $K_i$. 
For  $q=2^m$, each candidate code is a linear code with a parity-check matrix 
of size $m\times mK_i$. In other words, a parity-check matrix of the constituent code represents concatenation of $K_i$ companion matrices of the field elements.  
In the lists, constituent codes are represented by sorted sequences of $K_i$  degrees of the field primitive element. Random permutations of the field elements in the sequences  are taken into account in the process of the overall matrix optimization. To avoid the search over equivalent codes, the first sequence element is always equal to 1. 

For small values of $m$ and low rate codes (small value of $K_i$), the list of candidate codes can be obtained by the
exhaustive search. Otherwise, the codes are selected at random. In both cases, the candidate 
selection criterion is the minimum distance of the linear code. For the codes with 
the same minimum distance, we prefer codes with a smaller number of the minimum weight 
codewords.  

In the course of experimentation, for each value of $K_i$, the list of 50 code candidates was constructed.        
The algorithm for searching for the coefficient matrix $H_{c}$ is shown in Fig. \ref{Search}. 
The procedure uses a base matrix found by the simulated annealing 
technique and a degree matrix optimized by a greedy search in~\cite{Boch2016}.
The iterative procedure for labeling of the resulting base matrix by the field elements 
consists of random assigning of the good constituent code-candidates to the rows 
of the parity-check matrix and testing of the generalized girth of the resulting 
code. The newly generated code is considered the best one if it has either 
larger generalized girth, or if it has the same girth and lower 
multiplicity of the shortest cycles. The search stops if during the last $I_{\max}$ attempts,  
a new record was not achieved.   
 

%

\begin{figure}
\begin{algorithmic}[0]
\Statex{{\bf Input:} Base matrix $B$, degree matrix $H_w$, lists of candidate constituent codes 
$A_i\subset {\rm GF}(2^m)^{K_i}$, $i=1,...,c-b$.}
\Statex {{\bf Initialization:} Set $I=0$, generalized 
girth $g=0$,  and  multiplicity of length $g$ cycles $N=0$;  }
\While {$I\le I_{\max}$ }
{\\
\hspace{5mm} {$I\gets I+1$;} 
\hspace{5mm} \For {$i=1$ to {$c-b$}} 
{\\
\hspace{10mm} Choose at random a constituent code $\vec a =(a_1, ...,a_w)$ from \\
\hspace{10mm}  the list $A_i$
and permute it randomly 
$\vec a'={\rm randpermut}(\vec a)$;\\
\hspace{10mm} Assign components of  $\vec a'$ to nonzero elements of the 
$i$th row \\ \hspace{10mm}  of the matrix of coefficients $\tilde {H}_c$. 
}\EndFor\\
\hspace{5mm}Compute new generalized girth  value $\tilde {g}$ and multiplicity $\tilde {N}$ of 
length $\tilde {g}$ cycles;  
\hspace{5mm}\If { ($\tilde {g}>g$) $||$ (($\tilde {g}==g $) \&\& ($\tilde {N}<N$))}\\
\hspace{10mm} set $g=\tilde {g}$, $N=\tilde{N}$, $H_c=\tilde {H}_c$, $I=0$;
\EndIf
}
\EndWhile
\Statex{{\bf Output}: Matrix $H_c$}
\end{algorithmic}
\caption{Algorithm for searching for the coefficient matrix  $H_{c}$\label{Search}
}
\end{figure}

\section{Bounds on error probability \label{AlmostRegular}} 

In this section, we compute the tangential-sphere (TS) upper bound  \cite{poltyrev1994bounds} on the error probability of the maximum-likelihood (ML) decoding  for a random ensemble of ``almost regular'' NB LDPC codes.
In the appendix, for the sake of completeness of the paper, we present the upper bound as well as the Shannon lower bound on the error probability of ML decoding.

\subsection{Computing spectra of the ensembles of almost regular NB LDPC codes }

It is easy to see that in order to compute the TS
bound (\ref{TSB}) -- (\ref{TSB_eq}), it is necessary to know the weight spectrum of the code.  
In the sequel, we compute the average binary weight spectrum of an ensemble of NB LDPC codes over GF$(2^m)$, $m>1$ is an integer, with two and three nonzero elements in each column of their parity-check matrix. We start with a brief overview of ensembles of LDPC codes studied in literature. Then, the proposed ensemble is described and its average binary weight spectrum is derived.

\subsection{Ensembles of LDPC codes}   

Asymptotic distance spectra of ensembles of both regular and irregular binary LDPC codes were studied in \cite{burshtein2004asymptotic}. The main idea behind the approach in~\cite{burshtein2004asymptotic} is that it is possible to represent an LDPC code by its Tanner graph and replace the analysis of the ensemble of irregular LDPC codes by the analysis of the ensemble of irregular bipartite graphs with a given degree distributions on variable and check nodes, 
$\lambda(x)=\sum_{i=1}^{J}\lambda_ix^{i-1}$  and $\rho(x)=\sum_{i=1}^{K}\rho_ix^{i-1}$, where $\lambda_i$ and $\rho_i$ are a fraction of nodes of degree $i$ among the variable and check nodes, respectively, and 
 $J$ and $K$ are the maximum column and row weights. To each variable node of degree $i$, we assign $i$ variable edges, and to each check node  of degree $i$, we assign $i$ check edges, respectively, where $1\le i \le |\mathcal E|$ and $|\mathcal E|$ denotes the total number of edges in the graph.

In \cite{burshtein2004asymptotic},
an ensemble of random graphs with $n$ variable and $r$ check nodes is generated 
by assigning $J_i$ edges to each of the $\lambda_i n$ variable nodes. The edges are taken from the set $\mathcal E$, $i=1,2,...,K$. Next, all the edges are randomly permuted
by choosing uniformly at random a permutation $\bs \pi=(\pi_1,\pi_2,...,\pi_{|\mathcal E|})$ of the set $\{1,2,...,|\mathcal E|\}$. 
Then, each of the $\rho_j r$ check nodes is connected with $K_j$ edges from the permuted set of edges, $j=1,2,...,J$.

To map the corresponding graph to the code matrix $H$, the entry $H_{ij}$ is set to 1 if there is an odd number of edges between the $j$th variable node and the $i$th check node. Otherwise, $H_{ij}$ is set to 0. Notice that the described ensemble was introduced in  \cite{luby2001efficient}.

Another ensemble of irregular binary LDPC codes was considered in \cite{litsyn2003distance}. Binary LDPC codes in this ensemble are given by their random parity-check matrices having the following properties. The rows of the $r\times n$ parity-check matrix are split into $g$ horizontal strips, where the $i$th strip  contains $r\nu_i$ rows, $\sum_{i=1}^g \nu_i=1$. Its columns are spit into $h$ blocks, where the $i$-th block contains $n\eta_i$ columns,  $\sum_{i=1}^h \eta_i=1$. 
The sum of the elements in each row in the $i$th strip, $i=1,2,...,g$, is equal to $r_i$, and the sum of the elements in each column in the $i$th block,
is equal to $s_i$, where $r_1,...,r_g$ and $s_1,...,s_h$ are nonnegative integers independent of $n$. 
The required column and row degree distributions for an irregular code can be obtained by a proper choice of parameters $g$, $h$, and 
sequences $\nu_i$,~$\eta_i$,~$s_i$,~$r_i$.

Any ensemble of binary LDPC codes can be straightforwardly generalized to the ensemble of NB LDPC codes by randomly assigning elements of GF$(q)$ to all nonzero elements in the parity-check matrix.  
Similarly to \cite{burshtein2004asymptotic}, an ensemble of NB LDPC codes over GF$(2^m)$, which is determined by the ensemble of irregular bipartite graphs with a given degree distributions on variable and check nodes, and where each edge is labeled by an element in GF$(2^m)$, was studied in \cite{ kasai2011weight}. In particular, the average symbol weight and bit weight spectra of the random ensemble of irregular NB LDPC codes were derived therein. 

For the finite-length analysis both the ensemble obtained from the random bipartite graphs in 
\cite{luby2001efficient} and its generalization to the NB case in \cite{kasai2011weight} have the same shortcoming.
They do not give irregular codes with predetermined column and row weight distributions    
 $\lambda(x)$ and $\rho(x)$. Due to unavoidable parallel edges in the Tanner graph, 
the true degree distributions differ from the expected, and this phenomenon complicates the finite-length
analysis of the ensemble. Finite-length analysis for the ensemble in \cite{litsyn2003distance} is even more challenging. Only asymptotic
generating functions for the code spectra were found in \cite{litsyn2003distance}.

Ensembles of both binary and NB regular LDPC codes were first analyzed by Gallager in \cite{gallager}. Later, several ensembles of binary LDPC codes were studied in \cite{litsyn2002ensembles}. The average weight spectra for the corresponding  ensembles of regular LDPC codes were derived in  \cite{gallager} and \cite{litsyn2002ensembles}. In \cite{andryanova2009binary}, the asymptotic average weight spectra for ensembles of regular NB LDPC codes over GF$(2^m)$ were obtained. In \cite{bocharova2017}, we presented a low-complexity recurrent procedure for computing the exact spectra of both binary and NB random ensembles of regular  LDPC codes.

In the current paper, we deal with codes having only two and three nonzero elements in each column of their parity-check matrices. Therefore, we slightly modify the ensemble of regular NB LDPC codes and the corresponding low-complexity procedure for computing the average spectra in \cite{bocharova2017}. In the next subsection, we describe the new ensemble of ``almost regular'' NB LDPC codes over GF$(2^m)$ and present a generalized procedure to computing its average weight spectra. 

\subsection{Average binary weight spectrum of the ensemble of almost regular NB LDPC codes over GF$(2^m)$ \label{spec}}

The binary weight distribution of a linear code from a random ensemble can be represented via its weight generating function
\[G_n(s)=\sum_{w=0}^nA_{n,w}s^w,\]
where $A_{n,w}$ is a random variable representing the number of binary words of weight $w$ and length $n$. We aim at computing $\rm E\{A_{n,w}\}$, 
where $\rm E\{\cdot\}$ denotes the mathematical expectation over the code ensemble.
Next, we describe a new ensemble of NB LDPC codes over GF$(2^m)$ and derive its weight generating function. This ensemble can be viewed as a modification of the Gallager ensemble of $q=2^m$-ary LDPC codes.  
  
 In the well-studied Gallager ensemble  of binary 
 $(J,K)$-regular codes, the parity-check matrix of rate $1-r/n$ consists of $J$ strips
$H_{b}^{\rm T} = \begin{pmatrix}
H_1^{\rm T} \, | \, H_2^{\rm T} \, | \, \ldots \, | \, H_J^{\rm T}
\end{pmatrix}^{\rm T}$, where each strip  $H_i$ of width $M=r/J$
is a random permutation of the first strip  which can be chosen in the form 
\[
H_1=(\underbrace{    I_{M}\;...\;I_{M}\;}_{ K}),
\] 
$I_{M}$ is the identity matrix of order $M$.

We generalize the Gallager ensemble of binary LDPC codes by combining a given number  $K_i\le K$ 
of identity matrices and $K-K_i$ of all-zero $M\times M$ submatrices in the strip. Without loss of generality, the $i$th strip  can be chosen 
as a random permutation $\pi_i (\tilde  H_i) $, where $\tilde  H_i$ has the form:
\[
\tilde  H_i=(\underbrace{    I_{M}\;...\;I_{M}\;}_{ K_i} \underbrace{ \bs 0_M\;...\;\bs 0_M }_{K-K_i} )\;, \quad i=1,...,J,
\] 
and $\bs 0_M$ is the all-zero matrix of order $M$.
In this case, the strips  in the generalized ensemble are permuted versions of the strips  in the Gallager ensemble with 
some of the identity matrices replaced by the all-zero matrices of the same order.
By choosing the value of $K_i$, we could adjust the column weight and row weight distributions.  
In what follows, we will be mostly interested in parity-check matrices with $J=3$, that is, the column weights are all equal to two or three. 


An example of the parity-check matrix for the rate 1/2 LDPC code in this ensemble is as follows: 
 
\begin{equation}
H_{b}=
\left(
\begin{array}{llllllllllll}
1&0|&1&0|&1&0|&1&0|&1&0|&0&0\\
0&1|&0&1|&0&1|&0&1|&0&1|&0&0\\ \hline
0&1&1&0&0&0&1&1&0&1&1&0 \\
1&0&0&1&1&1& 0&0& 1&0& 0&1\\\hline
0&1&1&0&1&0&0&1&1&0&0&1\\
1&0&0&1&0&1&1&0&0&1&1&0\\
\end{array}
\right),
\label{G_diag}
\end{equation}
where $K=6$, $J=3$,  $M=2$, $K_1=5$. The matrix consists of two weight-2 and ten weight-3 columns.

By following the approach in \cite{gallager}, we can write down the generating function of 
the number of binary sequences $\bs x$ of weight $w$ and length $n$ satisfying the equality $\bs x H^{\rm T}_i=\bs 0$, $i=1,2,...,J$:
 
\begin{equation}
G_{i}(s)=\sum_{w=0}^{n}G_{i, n,w}s^{w}= g_{i}(s)^{M}, \quad i=1,...,J,
\label{genf} 
\end{equation}
where
\[ 
g_i(s)=(1+s)^{K-K_{i}}\sum_{j=0}^{K_i}g_{ij}s^j=(1+s)^{K-K_{i}}\left((1+s)^{K_i}+(1-s)^{K_i}\right)/2 \; , 
\] 
$g_{ij}=\binom{K_i}{j}$ if $j$ is even, and $g_{ij}=0$ otherwise.     

The probability that the binary sequence $\bs x$ of weight $w$ and length $n$ satisfies $\bs xH_{i}^{\rm T}=\bs 0$ can be expressed as:
\begin{eqnarray}
p_i(w)&=&\frac{G_{i,n,w}}{\binom{n}{w}}. \label{p1w} \\
 {\rm E}\{A_{n,w}\}&=&\binom{n}{w}^{1-J}\prod_{j=1}^{J} G_{j,n,w}\label{av_spec}.
 \end{eqnarray}

Consider the same generalization of the Gallager ensemble of the $q$-ary LDPC codes, where $q=2^m$, $m\ge 1$ is an integer. Specifically, in the $i$th strip of the parity-check matrix in the Gallager ensemble of the $(J,K)$-regular $q$-ary LDPC codes, we allow $K-K_i$ submatrices of order $M$ to be all-zero. The weight generating function of the $q$-ary sequences $\bs x$ of length $n$ satisfying the
nonzero part of one $q$-ary parity-check equation can be easily obtained by modifying the generating function in \cite{gallager}. It has the form 
 \[
f_{j}(s)=(1 + (q-1)s)^{K-K_{j}}  \frac{\left(1+(q-1)s\right)^{K_{j}}+(q-1)(1-s)^{K_j}}{q} \; .
\]
By assuming binomial probability distribution of zeros and ones in the $m$-dimensional binary image of the $q$-ary symbol, we can write down the average binary weight generating function for the $j$-th strip  as 
\begin{equation}
F_{j}(s)=\sum_{w=0}^{nm}F_{j, nm,w}s^{w}=f_{j}(\phi(s))^{M}
%
\label{genfNB} \; , 
\end{equation}
where
$F_{j, nm,w}$ denotes the average number of binary sequences $\bs b$ of weight $w$ and length $nm$ satisfying $\bs bB_{j}^{\rm T}=\bs 0$, $B_{j}$ is a binary image  of $H_j$, and 
\vspace{-2mm}
\begin{eqnarray}
\phi(s)&=&\sum_{i=1}^{m}\frac{1}{q-1}\binom{m}{i}s^{i}=\frac{(1+s)^{m}-1}{q-1} \; .
\end{eqnarray}
Analogous to (\ref{p1w}) and (\ref{av_spec}), we obtain 
\begin{eqnarray}
p_j(w)&=& \frac{F_{j,nm,w}}{\binom{nm}{w}} \label{pqw} \; , \\
{\rm E}\{A_{n,w,m}\}&=&
\binom{nm}{w}
\prod_{j=1}^{J} p_j(w) \; = \; \binom{nm}{w}^{1-J}\prod_{j=1}^{J}F_{j,nm,w}\; .
 \label{av_specNB}
 \end{eqnarray}
Thus, the problem of computing finite-length average spectra of NB LDPC codes is reduced to a problem of computing 
coefficients of the series expansion of the functions $F_j(s)$ in (\ref{genfNB}). This can be achieved
either directly by multiplying the polynomials, or recursively as in \cite{bocharova2017FL,bocharova2017}. 
In both cases,  the accuracy of the computations could be improved by working in the logarithmic domain.

\section{Simulations and comparisons \label{Simul}} 

In this section, first we present the tightened upper bounds on the frame error probability of the maximum-likelihood decoding  for finite length NB LDPC codes in the random ensemble described in Section \ref{spec}. These bounds are obtained by substituting the average binary weight spectra 
(\ref{av_specNB}) into the Poltyrev bound (\ref{TSB}). Comparison of these bounds with the Shannon lower bound and the Poltyrev bound for the random binary linear code  of the same length is performed. In the sequel, we use notation SNRb for the signal-to-noise ratio per bit measured in dB, $w$ is the average column weight, $J$ and $K$ denote the maximal number of nonzero elements in each column and each row of the parity-check matrix, respectively.    

We consider the rate $R=1/2$ NB LDPC codes with the maximum column weight in their parity-check matrices equal $J=3$.
The corresponding ensembles of the random almost regular NB LDPC codes are determined by the $3 \times 6$ base matrix. In all the examples, 
the parity-check matrices have row weights $K_1=K-\alpha, K_2=K_3=K$. 
The parameter $\alpha$ characterizes sparseness  of the parity-check matrices. The values $\alpha=0,1,2,3$ correspond to the average 
column weights $w=3$, $w=17/6=2.83$, $w=16/6=2.67$, and $w=15/6=2.50$, respectively.

It is expected that the codes with sparser parity-check matrices have higher error rate under the ML decoding. However, the
sparseness is important for performance improvement under the BP decoding. In the sequel, we optimize the sparsity 
of the parity-check matrices, which provides for good trade-offs between the error rates 
of the ML decoding and of the BP decoding.

\begin{figure}
\begin{center}
\includegraphics[width=95mm]{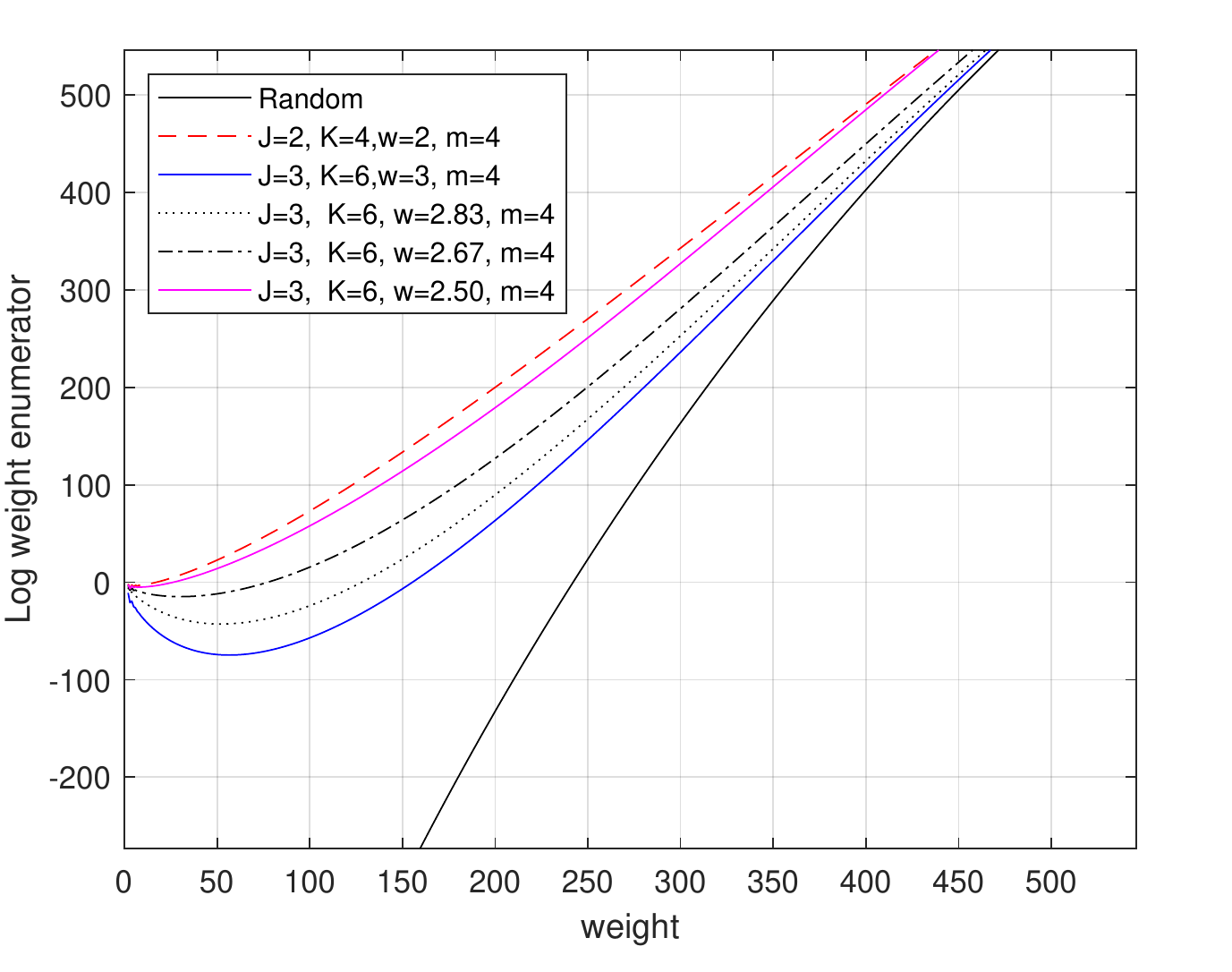}   
\caption{\label{sm4} The average binary weight spectra of rate $R=1/2$ NB QC LDPC codes of length 2080 bits over GF$(2^4)$ }
\end{center}
\end{figure}

\begin{figure}
\begin{center}
\includegraphics[width=95mm]{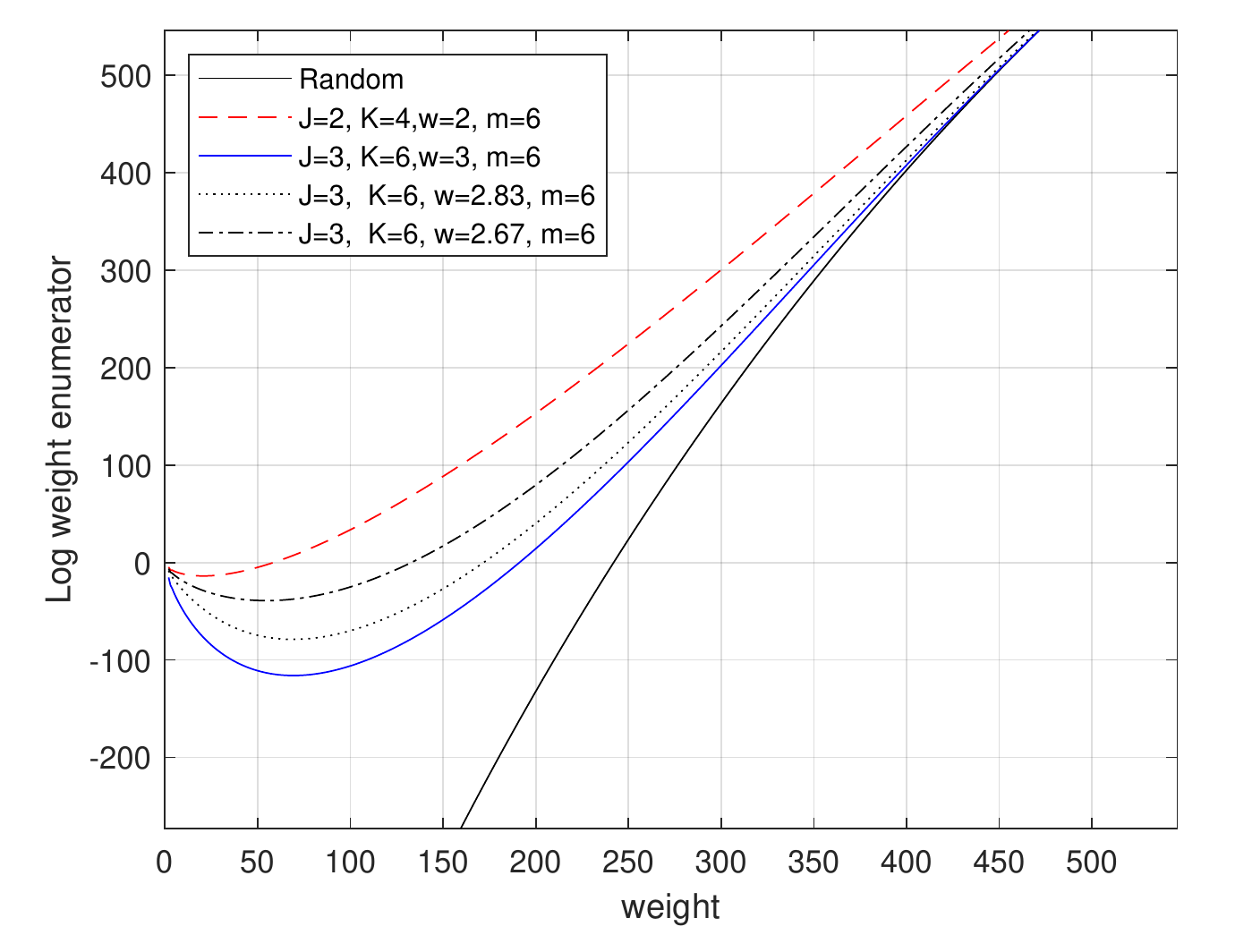}   
\caption{\label{sm6} The average binary weight spectra of rate $R=1/2$ NB QC LDPC codes of length 2080 bits over GF$(2^6)$ }
\end{center}
\end{figure}

The average binary weight  spectra of the ensembles of the NB LDPC codes of length about 2000  bits 
with various average column weights $w$ of their parity-check matrices are shown in Figs. \ref{sm4}--\ref{sm6}.

In particular, in Fig.~\ref{sm4} we show the average binary weight spectra for the random NB LDPC codes over GF$(2^4)$. In Fig.~\ref{sm6}, the average binary weight spectra for the random NB LDPC codes over GF$(2^6)$ are shown.

The average binary weight  spectra of the NB LDPC codes for $m=6$ are closer to the random linear binary code spectra than those for $m=4$ . Moreover, 
 if $m=6$ then even the ensemble of the (2,4)-regular NB LDPC codes has rather large average minimum distance (close to 50), which makes it potentially efficient.  It follows from the plots that denser parity-check matrices
are needed in order to achieve a near-optimal performance for $m=4$ than for $m=6$.   

\begin{figure}
\begin{center}
\includegraphics[width=95mm]{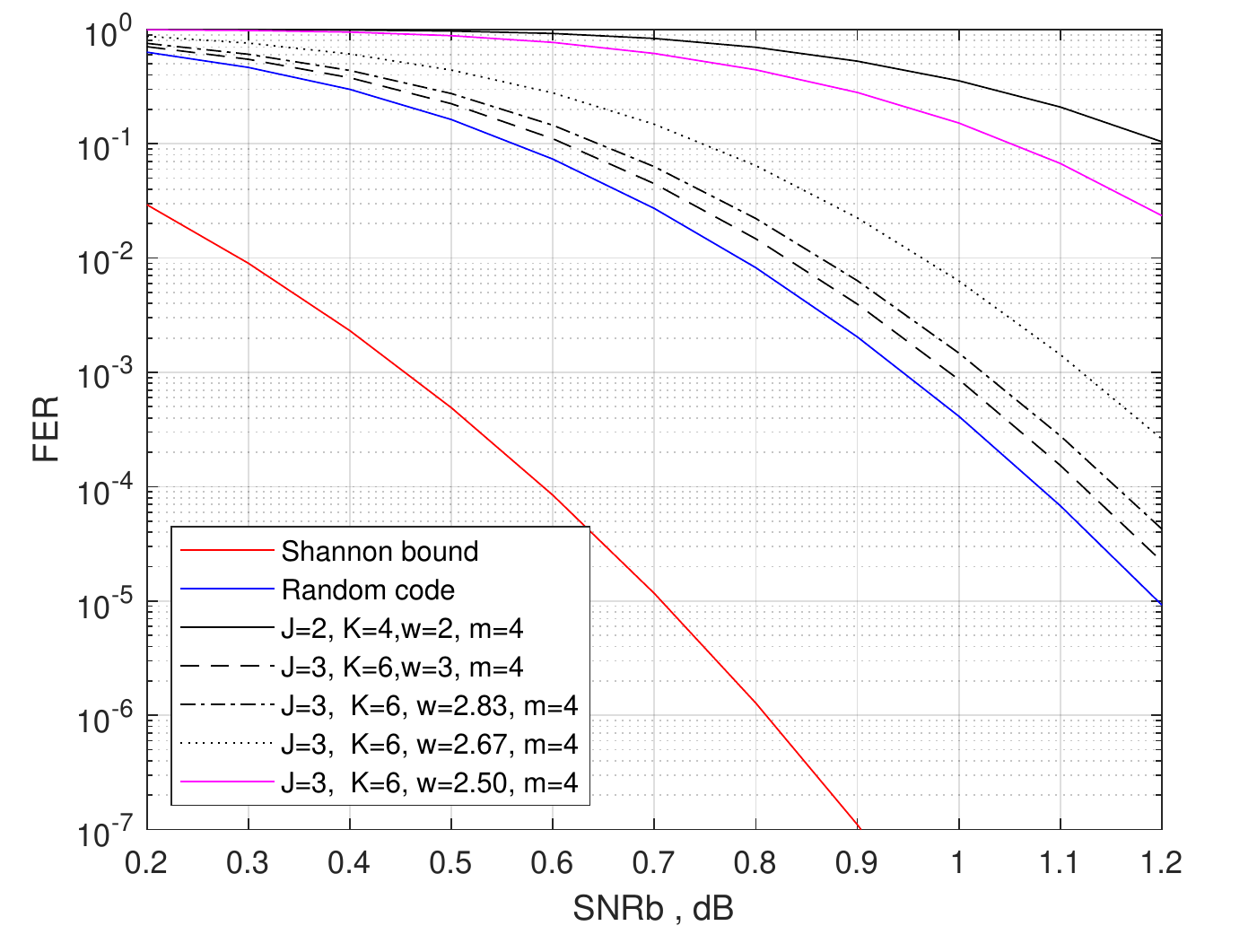}   
\caption{\label{bm4} The Poltyrev bounds on  the ML frame error probability of rate $R=1/2$ random NB LDPC codes of length 2080 bits over GF$(2^4)$ }
\end{center}
\end{figure}

\begin{figure}
\begin{center}
\includegraphics[width=95mm]{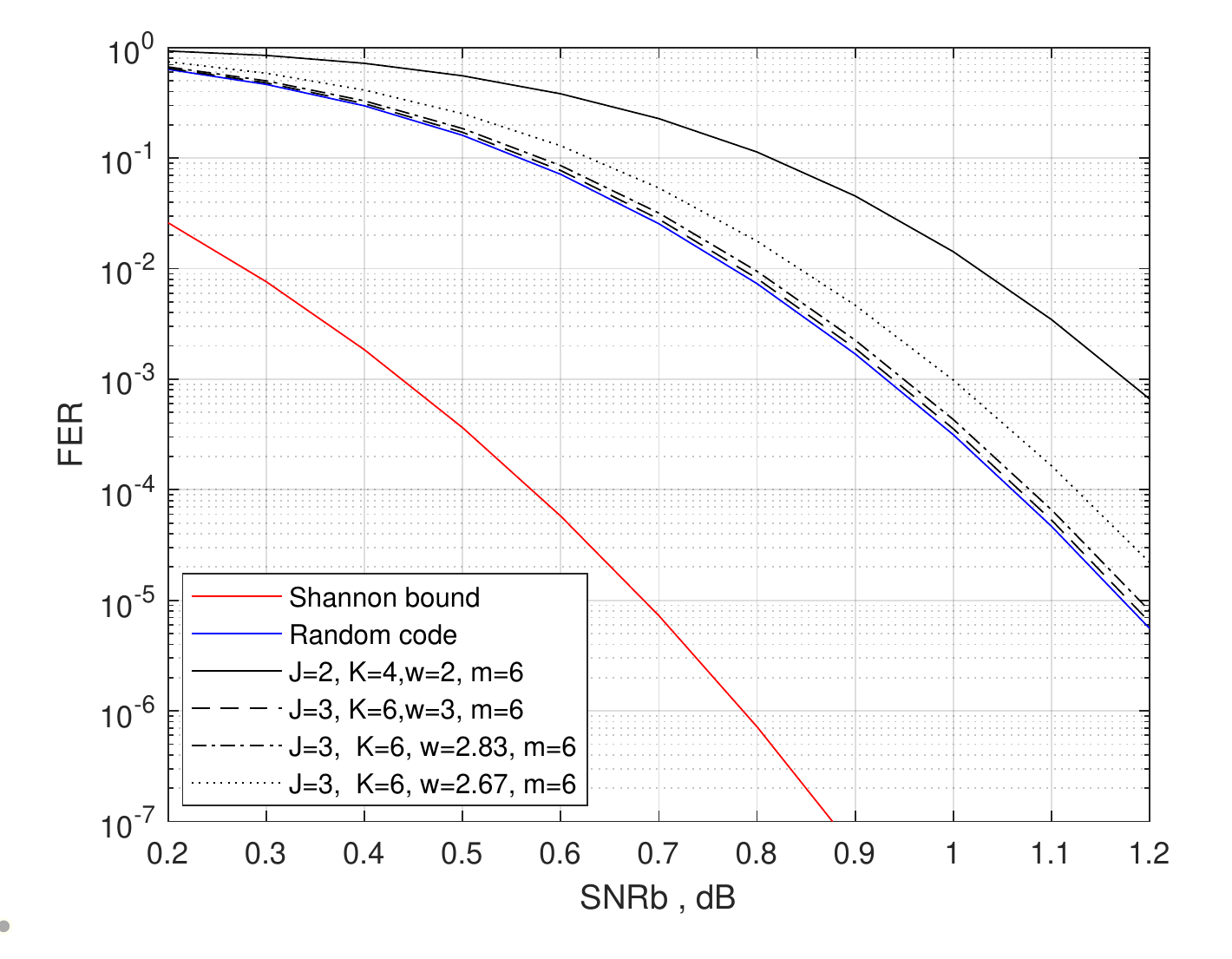}   
\caption{\label{bm6} The Poltyrev bounds on  the ML frame error probability  of rate $R=1/2$ random NB LDPC codes of length 2184 bits over GF$(2^6)$ }
\end{center}
\end{figure}

The corresponding random coding upper bounds on the frame error probability of the ML decoding over the AWGN channel are presented in Figs.~\ref{bm4}--\ref{bm6} for $m=4$ and $m=6$, respectively.
For comparison, the Poltyrev bound on the maximum-likelihood decoding frame error probability of the random linear binary codes  and the lower Shannon bound are shown in the same figures.

It follows from the presented plots that:
\begin{itemize}
\item{For $m=6$, the random NB LDPC code with only two nonzero elements in each column of its parity-check matrix loses about 0.2 dB in SNRb compared to the random linear code. However, for $m=4$, the corresponding loss in the performance is larger than or equal to 0.6 dB.}
\item{For $m=6$, the bound on the FER performance of the ML decoding for the random $(3,6)$-regular NB LDPC code coincides with the same bound for the random linear binary code. If $m=4$, we observe a small gap in the performance of the corresponding codes.} 
\item{For $m=6$, reduction in the column weight influences the ML decoding performance to a smaller extent than in the case $m=4$.}
\end{itemize}
 
Next, we compare the frame error rate (FER) performance of the sum-product BP decoding of rate $R=1/2$  NB QC-LDPC codes of length about 2000 bits with different average column weight $w$ in their base parity-check matrices observed in the simulations   
with the performance of the rate $R=1/2$ standard binary QC-LDPC code of length 2096 bits from the 5G standard. We consider almost regular NB QC-LDPC codes determined by the base matrix of size $26\times 52$ with the column weights two and three. The lifting  factor $L$ is chosen to be equal to 10, 8, and 7 for the codes over $2^4$, $2^5$, and $2^6$, respectively. The average column weight $w$ takes on values from the set $\{2.0, 2.19, 2.29,2.38\}$. The simulations were performed until twenty frame errors were encountered. 

\begin{figure}
\begin{center}
\includegraphics[width=95mm]{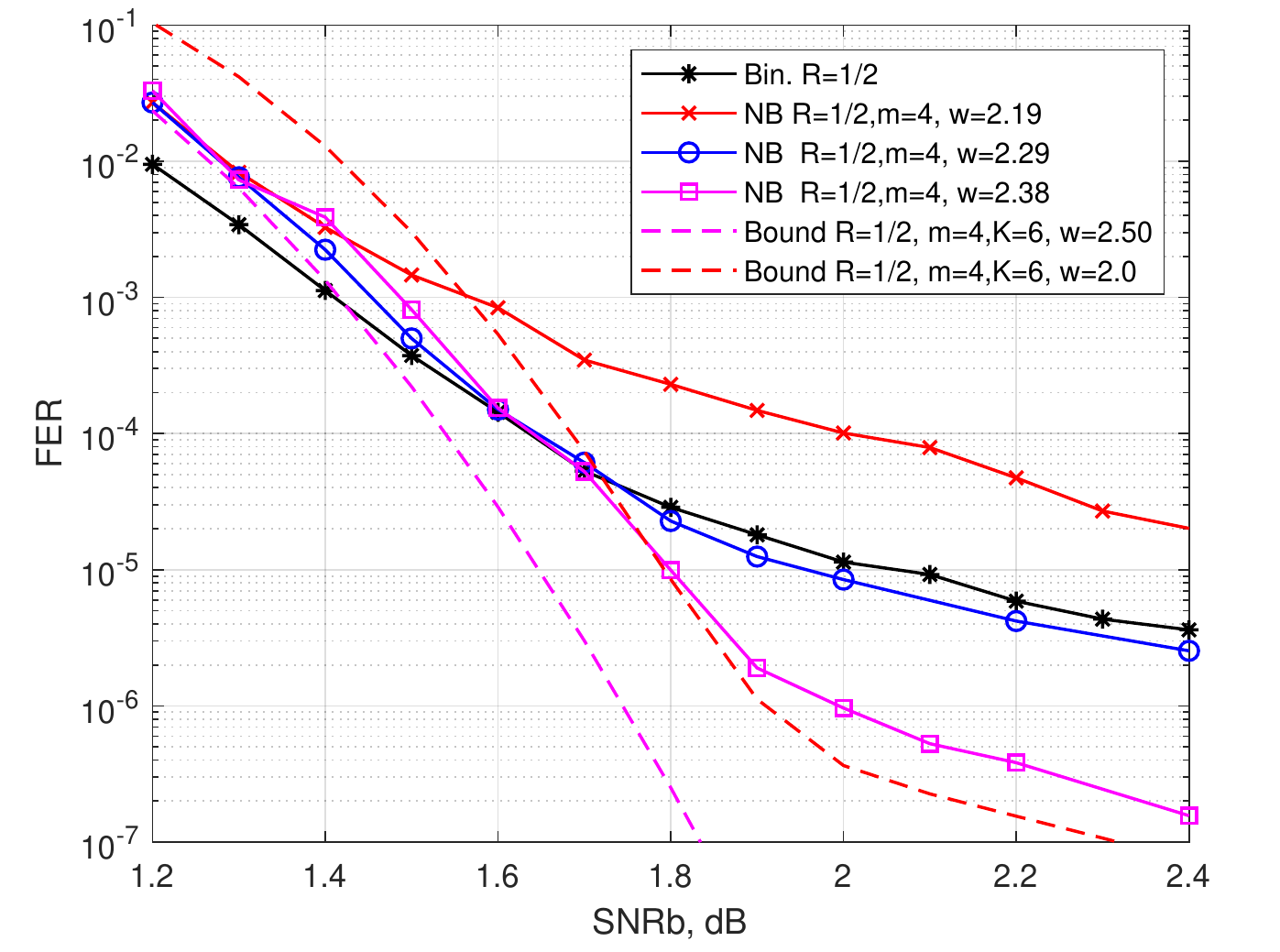}   
\caption{\label{fm4} FER performance of rate $R=1/2$ NB QC LDPC codes
over GF($2^4$) 
of binary length 2080 bits }
\end{center}
\end{figure}

\begin{figure}
\begin{center}
\includegraphics[width=95mm]{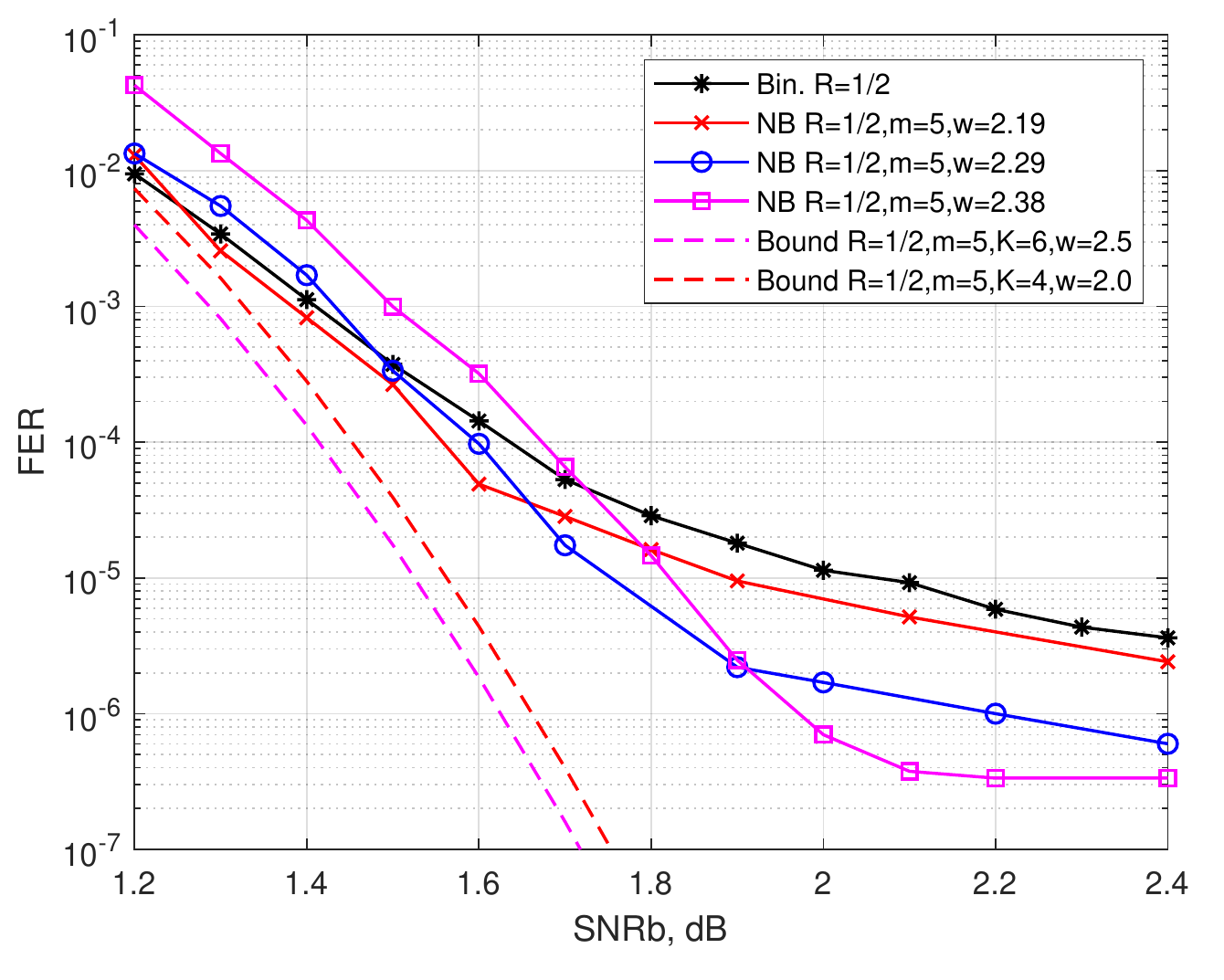}   
\caption{\label{fm5} FER performance of rate $R=1/2$ NB QC LDPC codes 
over GF($2^5$) 
of binary length 2080 bits }
\end{center}
\end{figure}

\begin{figure}
\begin{center}
\includegraphics[width=95mm]{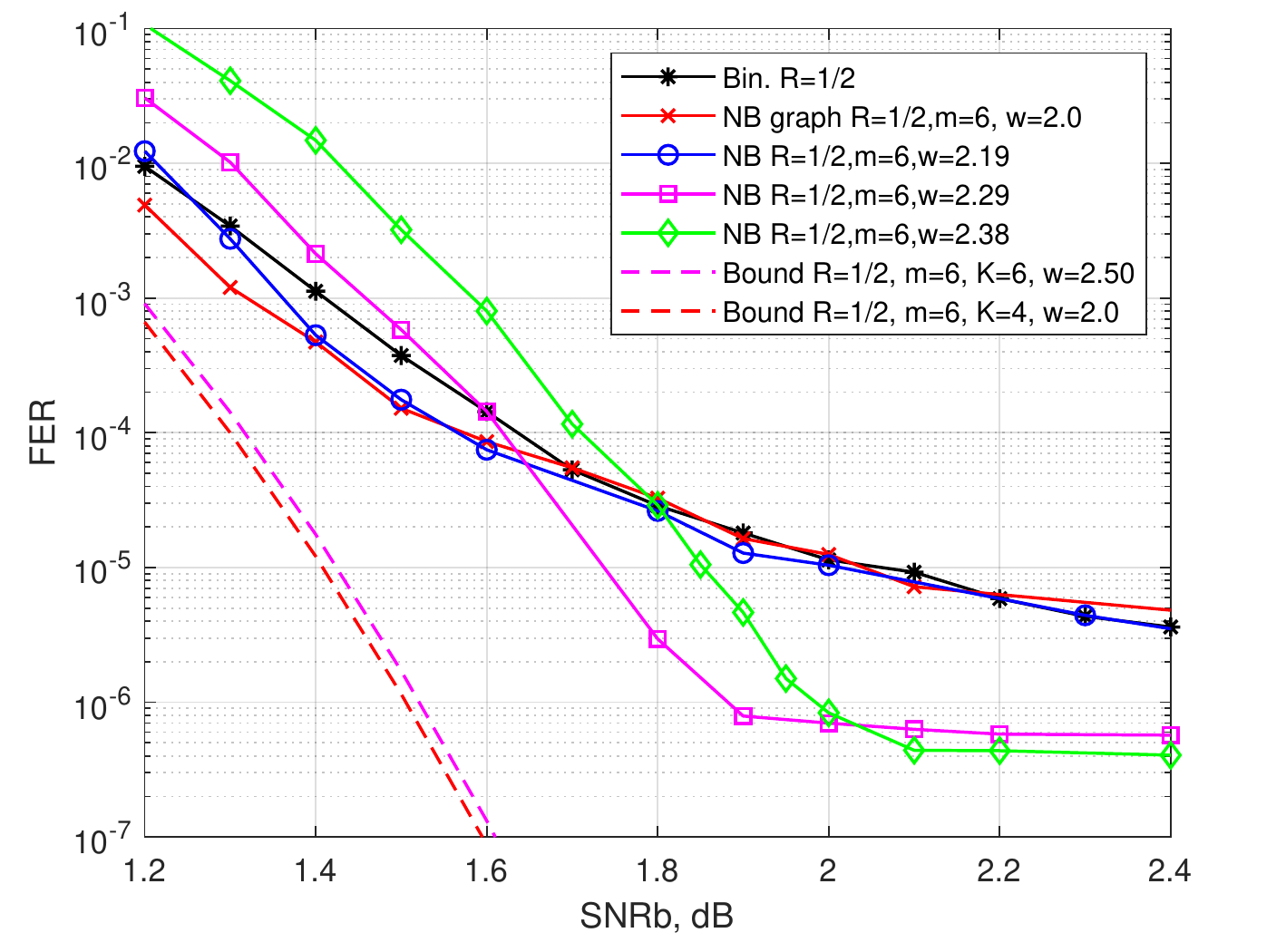}   
\caption{\label{fm6} FER performance of rate $R=1/2$ NB QC LDPC codes 
over GF($2^6$) of binary length 2184 bits }
\end{center}
\end{figure}

The corresponding plots are shown in Figs. \ref{fm4}--\ref{fm6}.

 From the presented plots, we can conclude the following:
 \begin{itemize}
 \item
If the field extension degree  is $m=4$, then in the low SNRb region, the NB QC-LDPC codes are inferior to the optimized binary QC-LDPC code of the same length. Moreover, the NB QC LDPC code having average column weight $w=2.19$ in its base parity-check matrix is inferior to the optimized binary QC-LDPC code in the entire SNRb region. The increase in $w$ improves the FER performance almost monotonically. The NB code with $w=2.38$ gains at least 0.6 dB in the high SNRb region when compared to the binary code.        
 \item 
When $m$ increases, we observe that the FER performance 
in the error floor region improves monotonically with increase in $w$.
In the waterfall region, there exists an optimal value of $w$, which provides for the best performance. For $m=5$, the optimal value is $w=2.19$. If $m=6$, then the optimal value is $w=2.0$.
\item
The gap between the theoretical bound and the simulation results is due to the two factors: imperfectness 
of the constructed codes and suboptimality of the BP decoding algorithm. 
For all the codes in the simulation we observe the error floor. 
We conjecture that the class of QC LDPC codes has 
limited achievable performance.
\item
Binary codes in the 5G standard demonstrate very good FER performance uner the BP decoding in the waterfall region, where they compete with the NB QC-LDPC codes.  However, in the error floor region, the performance of the NB LDPC codes is superior to that of their binary counterparts.
\item
The presented simulation results in the waterfall region are about 0.1~--~0.2 dB away from the tightened random coding bounds on the FER performance of the ML decoding. 
\end{itemize}

\section{Conclusion}
A new optimization technique for constructing NB QC-LDPC codes was proposed and analyzed. The key feature of the new technique is that it is based on the simulated annealing approach to optimization of the base parity-check matrices of the NB QC-LDPC codes.

The new ensemble of irregular NB LDPC codes was introduced and analyzed. The similarity of this ensemble to the Gallager ensemble of regular LDPC codes allowed us to use an important advantage of the Gallager ensemble, namely, the simplicity of its analysis. By substituting the computed average binary spectra for this new ensemble into the Poltyrev upper bound, the tightened finite-length upper bounds on the error probability of the ML decoding for irregular NB LDPC codes over GF($2^m$) were derived.  

The presented simulation results and comparisons with the bounds on error probability suggest that the NB QC-LDPC codes outperform the known binary QC-LDPC codes but as their binary counterparts suffer from severe error floor. Thus, further improvements on both the optimization method and the decoding algorithms for this class of codes can be considered as subject of the future research.

\begin{appendix}
\subsection{Lower bound}
In the sequel, we use approximation \cite{wiechman2008improved} 
of the Shannon lower bound \cite{Shannon1959}.

Denote by $n$, $R$, and $\sigma$ the code length, code rate and standard noise deviation for an AWGN channel, respectively. We use the notations and expressions in \cite{Shannon1959} for the cone half-angle $\theta\in [0,\pi]$, which
corresponds to the solid angle of an $n$-dimensional circular cone, and for the solid angle of the whole space
\[
\Omega_{n}(\theta)=\frac{2\pi^\frac{n-1}{2}}{\Gamma(\frac{n-1}{2})}\int_{0}^{\theta}(\sin \phi)^{n-2}d\phi \; , 
\;\;\;
\Omega_{n}(\pi)=\frac{2\pi^{n/2}}{\Gamma(n/2)} \; ,
\]
respectively.
For a given code of length $n$ and cardinality $2^{nR}$, the parameter $\theta_0$ is selected as a solution of the equation
\[
\frac{\Omega_{n}(\theta_{0})}{\Omega_{n}(\pi)}=2^{-nR} \; .
\]
The  approximation~\cite[Theorem~4.2]{wiechman2008improved} 
for the Shannon
lower bound \cite{Shannon1959} on  the FER  $P_{\rm sh}(n,R,\sigma)$ is

\begin{equation}
P_{\rm sh}(n,R,\sigma) \ge
\frac{\sigma\sqrt{n-1}}{6n(1+\sigma)}
\exp\left\{\frac{3\sigma^2-(\sigma+1)^2}{2\sigma^2}   \right\}
e^{-nF_L(\theta)}\;,
\label{Shannon1}
\end{equation}
where 
\begin{eqnarray}
G(\theta)&=&\frac{1}{2\sigma}\left(\cos\theta+\sqrt{\cos^{2}\theta+4\sigma^2}\ \right)\;,\\
F_L(\theta)&=&\frac{1}{2\sigma^2}
\left(
1-\sigma G(\theta) \cos(\theta) -\sigma^2 \ln (G(\theta)\sin(\theta))
\right) \;.
\end{eqnarray}

\subsection{Upper bound}
The Poltyrev  bound \cite{poltyrev1994bounds} is the most tight TS-type bound:   
\begin{eqnarray}
P_e &\le&  \int_{-\infty}^{\sqrt n} f\left(\frac{x}{\sigma}\right)
\left\{
\sum_{w\le w_0} S_w \Theta_w(x)
\right.+
\nonumber\\
&+& \left. 1-\chi_{n-1}^2\left(\frac{r^2_x}{\sigma^2}\right)
\right\}dx + Q\left( \frac{\sqrt n}{\sigma}\right).
\label{TSB}
\end{eqnarray}
Here $f(x)=(1/\sqrt{2\pi})\exp{-x^2/2}$ is the  Gaussian probability  density function,  
$Q(x)  =  \int_{x}^{\infty} f(x) dx$, 
\begin{eqnarray*}
\Theta_w(x) & = & 
 \int_{\beta_w(x)}^{r_x}
f\left(\frac{y}{\sigma}\right)\chi_{n-2}^2
\left(\frac{r^2_x-y^2}{\sigma^2}\right)
dy \; , 
\\
w_0&=&\left \lfloor \frac{r_0^2n}{r_0^2+n}\right \rfloor ,
r_x=r_0\left(1-\frac{x}{\sqrt n}\right) \; ,\\
\mu_w(r)&=&\frac{1}{r}\sqrt {\frac{w}{1-w/n}}, 
\beta_w(x)=\left(1-\frac{x}{\sqrt n}\right)\sqrt {\frac{w}{1-w/n}} \; , 
\end{eqnarray*}
$S_{w}$ is the $w$-th spectrum coefficient,  and $\chi_{n}^2$ denotes the  probability density function of chi-squared distribution  with $n$ degrees of freedom.

Parameter $r_0$ is a solution with respect to $r$ of the equation
\begin{equation}\label{TSB_eq}
\sum_{w:\mu_w(r)<1}S_w \int_0^{\arccos \mu_w(r)} \sin^{n-3}\phi \quad d \phi =\sqrt{\pi} \cdot
\frac{\Gamma\left(\frac{n-2}{2}\right)}{\Gamma\left(\frac{n-1}{2}\right)} \; .
\end{equation}

 \end{appendix}
 
 \section*{Acknowledgment}

This work was supported by Huawei Technologies Co., Ltd. The authors wish to thank Victor Krachkovsky, Oleg Kurmaev, Alexey Mayevskiy and Hongchen Yu for helpful discussions.

\bibliographystyle{IEEEtran}
\bibliography{refer1510}
\end{document}